\documentclass[RNAAS]{aastex63}

\graphicspath{{./}}
\begin{document}

\title{Potential Backup Targets for \emph{Comet Interceptor}}

\author[0000-0003-4365-1455]{Megan E. Schwamb}
\correspondingauthor{Megan E. Schwamb}
\email{m.schwamb@qub.ac.uk}
\affiliation{Astrophysics Research Centre, School of Mathematics and Physics, Queen's University Belfast,Belfast BT7 1NN, UK}

\author[0000-0003-2781-6897]{Matthew M. Knight}
\affiliation{Department of Physics, United States Naval Academy, 572C Holloway Rd, Annapolis, MD 21402, USA}
\affiliation{University of Maryland, Department of Astronomy, College Park, MD 20742, USA}

\author[0000-0002-5859-1136]{Geraint H Jones}
\affiliation{Mullard Space Science Laboratory, University College London, Holmbury St. Mary, Dorking, Surrey, RH5 6NT, UK}
\affiliation{Centre for Planetary Sciences at UCL/Birkbeck, London, UK}

\author[0000-0001-9328-2905]{Colin Snodgrass}
\affiliation{Institute for Astronomy, University of Edinburgh, Royal Observatory, Edinburgh EH9 3HJ, UK}

\author{Lorenzo Bucci}
\affiliation{European Space Operations Centre, Robert-Bosch-Str. 5, 64293, Darmstadt, Germany}

\author{Jos\'{e} Manuel S\'{a}nchez P\'{e}rez}
\affiliation{European Space Operations Centre, Robert-Bosch-Str. 5, 64293, Darmstadt, Germany}

\author{Nikolai Skuppin}
\affiliation{European Space Operations Centre, Robert-Bosch-Str. 5, 64293, Darmstadt, Germany}

\collaboration{8}{for the \emph{Comet Interceptor} Science Team}

\section{\emph{Comet Interceptor} Mission} 

\emph{Comet Interceptor} \citep{2019NatCo..10.5418S} is an ESA (European Space Agency) F-class (Fast) mission expected to launch in 2028 on the same launcher as ESA's \emph{ARIEL} \citep[\emph{Atmospheric Remote-sensing Infrared Exoplanet Large-survey};][]{2016SPIE.9904E..1XT} mission. \emph{Comet Interceptor}'s science payload consists of three spacecraft, a primary spacecraft that will carry two smaller probes to be released at the target. 
The three spacecraft will fly-by the target along different chords, providing multiple simultaneous perspectives of the comet nucleus and its environment. Each of the spacecraft will be equipped with different but complementary instrument suites designed to study the far and near coma environment and surface of a comet or interstellar object (ISO), a rogue planetesimal formed around another star system passing through the Solar System 
The primary spacecraft will perform a fly-by at $\sim$1000 km from the target. The two smaller probes will travel deeper into the coma, closer to the nucleus. 

The mission is being designed and launched without a specific comet designated as its main target. \emph{Comet Interceptor} will travel to the Sun-Earth L2 Lagrangian point with \emph{ARIEL} and wait in hibernation until a suitable long-period comet (LPC) is found that will come close enough to the Sun for the spacecraft to maneuver to an encounter trajectory.  
Ideally, the mission will visit a dynamically new comet (DNC), a LPC on its first entry into the the inner Solar System from the Oort cloud, or an ISO. Since 1970, 
23$\%$ of LPCs discovered with q$<$3.1 au are DNCs \citep{2019MNRAS.484.3463K}. \emph{Comet Interceptor}'s encounter target will mostly likely be discovered in the 2020s with the onset of the Vera C. Rubin Observatory's Legacy Survey of Space and Time \citep[LSST;][]{2019ApJ...873..111I}. Over its 10-year baseline survey, LSST is expected to find $\sim$10,000 new comets \citep{2010Icar..205..605S} and to discover several ISOs per year  \citep[e.g.,][]{2016ApJ...825...51C,2019ApJ...884L..22R}. 

 \section{Request for Community Observations} 

In the event of a suitable LPC or ISO not being discovered before \emph{Comet Interceptor} is required to leave L2
, the science team will select a known short-period comet as the mission target. To prepare for all eventualities, the science team has assembled a preliminary set of backup targets from the known Jupiter family comets (see Table \ref{tab:backup_targets}), where a suitable fly-by trajectory can be achieved during 
the nominal mission timeline (including the possibility of some launch delay). To better prioritize this list, we are releasing our potential backup targets in order to solicit the planetary community's help with observations of these objects over future apparitions and to encourage publication of archival data on these objects.
Observations characterizing the nuclei of these comets (size, shape, rotation period) and their activity levels (especially around perihelion) would be particularly useful. Astrometry and improved orbits, especially for the more recently discovered or fragmented comets
would also be beneficial.
Any additional information that may assist scientific prioritization of these targets
would be welcomed by the \emph{Comet Interceptor} team.

\begin{deluxetable}{lllllllll}
\tablecolumns{6}
\tablecaption{\label{tab:backup_targets} Potential \emph{Comet Interceptor} Backup Targets}
\tablehead{\colhead{MPC}  &  \colhead{Orbital} &  \colhead{q} & \colhead{e} &  \colhead{i} & \colhead{Fly-by} & \colhead{Notes}  \vspace{-0.25cm}  \\
 \colhead{Designation}& \colhead{Period} &  \colhead{(au)} & \colhead{(deg)} & \colhead{(deg)} & \colhead{date} & \vspace{-0.25cm} \\
 & \colhead{(yrs)}  &   &  &   & \colhead{\footnotesize{DD-MM-YYYY}} & }
\startdata 
\tableline
P/2016 BA14 (PANSTARRS) &  5.25 & 1.01 & 0.666 & 18.92 & 20/04/2037 & May have recently split (from 252P)$^A$\\
\hline
7P/Pons-Winnecke  &  6.37 &1.26 &  0.634 & 22.29  & 28/09/2033 & \\
\hline
8P/Tuttle &  13.61 & 1.03 & 0.820 &  54.98  & 26/03/2035 & Contact binary; Oort cloud source?$^B$\\
\hline
15P/Finlay  &  6.51 & 0.97 & 0.720 & 6.80 & 18/09/2034 & \\
\hline
21P/Giacobini-Zinner &  6.54 & 1.01 & 0.710 & 32.00  & 07/09/2031 & Encountered by the \emph{International} \\ &   &  & &   & & \emph{Cometary Explorer} (\emph{ICE}; 1985)$^C$ \\
\hline
26P/Grigg-Skjellerup &  5.31 & 1.12 & 0.633 & 22.36  & 10/06/2034 & Encountered by \emph{Giotto} (1992)$^D$\\
\hline
73P/Schwassmann-Wachmann 3 &  5.44 & 0.97 & 0.686 & 11.24  & 05/04/2033 & Multiple nucleus fragments\\
\hline
189P/NEAT &   4.98 & 1.17 & 0.598 & 20.40  & 01/09/2032 & Possible escaped main belt asteroid$^E$ \\
\hline
289P/Blanpain &  5.32 & 0.96 & 0.685 & 5.90  & 23/11/2035 & Tiny, almost extinct$^F$\\
\hline
300P/Catalina &  4.42 & 0.82 & 0.694 & 5.70  & 19/06/2036 & Possible escaped main belt asteroid$^E$\\
\enddata 
\tablenotetext{A}{ \cite{2017AJ....154..136L}}
\tablenotetext{B}{ \cite{2010Icar..207..499H} }
\tablenotetext{C}{\cite{1986Sci...232..374O}}
\tablenotetext{D}{ \cite{1993JGR....9820907G}}
\tablenotetext{E}{  \cite{2015PSS..118...14F}}
\tablenotetext{F}{ \cite{2006AJ....131.2327J}}
\tablenotetext{}{Heliocentric orbital elements taken from JPL Small-Body Database Search Engine \url{https://ssd.jpl.nasa.gov/sbdb_query.cgi\#} queried on 28 January 2020. }
\end{deluxetable}

\bibliographystyle{aasjournal}
\bibliography{ms} 

\acknowledgments

 This work has made use of NASA's Astrophysics Data System Bibliographic Services. MES was supported by STFC Grant ST/P000304/1.

\end{document}